\documentclass[superscriptaddress,twocolumn,showpacs,amsmath,amssymb]{revtex4}
\usepackage{graphicx}
\usepackage{color}

\renewcommand{\vec}[1]{\mbox{\boldmath $#1$}}

\begin{document}

\title{
Decay dynamics of the unbound $^{25}$O and $^{26}$O nuclei}

\author{K. Hagino}
\affiliation{ 
Department of Physics, Tohoku University, Sendai 980-8578,  Japan} 
\affiliation{Research Center for Electron Photon Science, Tohoku University, 1-2-1 Mikamine, Sendai 982-0826, Japan}
\affiliation{
National Astronomical Observatory of Japan, 2-21-1 Osawa,
Mitaka, Tokyo 181-8588, Japan}

\author{H. Sagawa}
\affiliation{
RIKEN Nishina Center, Wako 351-0198, Japan}
\affiliation{
Center for Mathematics and Physics,  University of Aizu, 
Aizu-Wakamatsu, Fukushima 965-8560,  Japan}


\begin{abstract}
We study the ground and excited resonance states of $^{26}$O with a three-body model of 
$^{24}$O+n+n taking into account the coupling to the continuum.  
To this end, we use the new experimental data for the invariant mass spectroscopy 
of the unbound $^{25}$O and $^{26}$O nuclei, and present an update of 
three-body model calculations for the two-neutron decay of the $^{26}$O 
nucleus. 
With the new model inputs determined with the ground state decay of $^{26}$O, 
we discuss the di-neurtron correlations and a halo nature of this nucleus, as well as the structure of 
the excited states. 
For the energy of the 
2$^+$ state, 
we achieve an excellent 
agreement with the experimental data with this calculation. 
We show that the 2$^+$ state consists 
predominantly of the 
$(d_{3/2})^2$ configuration, for which the pairing interaction between 
the valence neutrons slightly decreases its energy from the unperturbed 
one. We also discuss the structure of excited 0$^+$ states of 
the $^{26}$O nucleus. 
In particular, we show the existence of an excited 
0$^+$ state at 3.38 MeV, which is mainly composed of the $(f_{7/2})^2$ 
configuration. 
\end{abstract}


\maketitle

\section{Introduction}

In recent years, 
there has been a rapidly increasing interest in 
two-neutron decays of unbound nuclei beyond the 
neutron drip line \cite{5H-1,5H-2,5H-3,5H-4,J10,J10-2,
S12,KSB12,SKB12,KLD13,LDK12,CSA13,KBB13,KBC15,Kondo15,A08,GMSZ11,GMZ13,GZ15,HS14,HS14-2,Kikuchi13}. 
These are similar phenomena to the two-proton radioactivities 
in unbound nuclei beyond the proton-drip line \cite{PKGR12}, but with neutrons. 
While the resonance in proton radioactivity is
 mainly due to the Coulomb barrier, the resonant 
two-neutron emission
arises from penetration of a centrifugal barrier.
Since the long range Coulomb interaction is absent in the two-neutron 
decays, it has in general been expected that nucleon correlations, such as 
the dineutron correlations \cite{MMS05,M06,HS05,SH15,PSS07}, 
are easier to be probed in the two-neutron 
decays as compared to the two-proton decays. 

Among the two-neutron emitters studied so far, the $^{26}$O nucleus has 
attracted a particular attention 
\cite{LDK12,CSA13,KBB13,KBC15,Kondo15,GMZ13,GZ15,HS14,HS14-2}, 
owing partly to the problem of abrupt termination of neutron-drip line 
for the oxygen isotopes at the neutron number 
$N$=16 \cite{S99,OSHSA10}. 
The ground state decay of this nucleus was first 
observed by Lunderberg {\it et al.} at the National Superconducting 
Cyclotron Laboratory (NSCL) at Michigan State University (MSU) 
\cite{LDK12}. 
A clear resonance peak was observed in the decay energy spectrum 
at $E=150^{+50}_{-150}$ keV \cite{LDK12}. 
This has been confirmed by the GSI-LAND group, who reported the 
upper limit of the decay energy to be 40 and 120 keV with 
the confident level of 68\% and 95\%, respectively \cite{CSA13}. 
These experimental data on the ground state decay of 
$^{26}$O have been theoretically analyzed in 
Refs. \cite{GMZ13,GZ15,HS14,HS14-2}. 

Very recently, new experimental data on the decay of the 
$^{25,26}$O nuclei came out from the radioactive ion beam factory (RIBF) at 
RIKEN, which have revised the previous data with 
much higher statistics \cite{Kondo15}. 
The energy of the ground state of $^{26}$O 
has now been determined with 
a higher precision to be 18 $\pm$ 3 (stat) $\pm$ 4 (sys) keV \cite{Kondo15}. 
Moreover, Ref. \cite{Kondo15} has also reported a clear second peak at 
1.28 $^{+0.11}_{-0.08}$ MeV \cite{Kondo15}, which is likely due to the 
excited 2$^+$ state. 
The data for the $^{25}$O have also been revised in this experiment. 
While the previous measurements reported 
the $d_{3/2}$ resonance 
state at 770$^{+20}_{-10}$ keV with the width of 172$\pm 30$ keV 
\cite{H08}, and at 725$^{+54}_{-29}$ keV with the width of 
20$^{+60}_{-20}$ keV \cite{CSA13}, the new measurement has shown 
the $d_{3/2}$ resonance 
state at 749 (10) keV with the width of 88 (6) keV \cite{Kondo15}. 

In this paper, 
we study the ground  and excited resonance states in $^{26}$O with a three-body model 
by taking into account the coupling to the continuum.  
The main aim of our study is to extract the di-neutron correlations and a halo nature 
of the ground state of $^{26}$O from the two-neutron decay spectrum 
with updated empirical inputs for the model Hamiltonian.  
In the present $^{24}$O +$n$ + $n$ three-body model,  the neutron-core potential 
as well as the strength of the pairing interaction between the valence neutrons 
are calibrated by the empirical data. 
To this end, 
we adopt  the new experimental data of Ref. \cite{Kondo15} and 
refine the calculations performed 
in Refs. \cite{HS14,HS14-2}.  
With the same model input,  we also discuss the structure 
of excited $0^+$ and $2^+$ resonance states, 
which were not presented in our previous publications. 

The paper is organized as follows. 
In Sec. II, we discuss the resonance structure of the $^{25}$O nucleus. 
We use the new experimental data for this nucleus to determine the $n$-$^{24}$O 
potential and obtain the resonance states of $^{25}$O. We also discuss how 
the Green's function can be used to estimate the width of the resonance states. 
In Sec. III, we discuss the decay energy spectrum for the $0^+$ configuration of the 
$^{26}$O nucleus. We also apply the bound state approximation and discuss 
the radius and the angular momentum configurations.  
In Sec. IV, we discuss the first 2$^+$ state, and make a comparison with other 
theoretical calculations. In Sec. V, we discuss the angular correlation of the emitted 
two neutrons and show that the back-to-back emission is enhanced due to the dineutron 
correlation. We then summarize the paper in Sec. VI. 

\section{Resonance structure of the $^{25}$O nucleus} 

\subsection{Calibration of the $n$-$^{24}$O potential and single-particle resonances}

An important input for the three-body calculation is the potential 
between a neutron and the core nucleus. 
In order to calibrate it, we first discuss the properties of 
the two-body subsystem of $^{26}$O, that is, the $^{25}$O nucleus, using the 
neutron + $^{24}$O model. 

Assuming that $^{24}$O is inert in 
the ground state, we consider the following single-particle Hamiltonian 
for the relative motion between a neutron and the core nucleus: 
\begin{equation}
h_{nC} = -\frac{\hbar^2}{2\mu}\vec{\nabla}^2+V_{nC}(r),
\label{Hsp}
\end{equation}
where $\mu=A_cm_N/(A_c+1)$ is the reduced mass, $m_N$ and $A_c=24$ 
being the nucleon mass and the mass number of the core nucleus, respectively. 
We employ the Woods-Saxon potential for the neutron-core potential, 
\begin{equation}
V_{nC}(r)=\left(V_0+V_{ls} (\vec{l}\cdot\vec{s})\frac{1}{r}\frac{d}{dr}\right)
\left[1+\exp\left(\frac{r-R}{a}\right)\right]^{-1},
\end{equation}
where $R=r_0A_c^{1/3}$. We use the same value for the diffuseness 
parameter, $a$, and the radius parameter, $r_0$, as in Ref. \cite{HS14}, 
that is, $a$=0.72 fm and $r_0$=1.25 fm. 
With these values of $a$ and $r_0$, the depth parameter, $V_0$, 
is determined to be $-44.1$ MeV in order to reproduce the energy of the 
2$s_{1/2}$ state, $\epsilon_{2s_{1/2}} = -4.09(13)$ MeV \cite{H08}. 
For the strength of the spin-orbit potential, $V_{ls}$, we use the new data 
for the energy of the unbound $d_{3/2}$ state, that is, 
$\epsilon_{d_{3/2}}$ = 749 (10) keV \cite{Kondo15}. 
To this end, we seek a Gamow resonance state by imposing the 
outgoing boundary 
condition to the radial wave function. 
The resultant value is $V_{ls}$ = 45.605 MeV fm$^2$, which is slightly smaller 
than the value employed in Ref. \cite{HS14}. 
This potential yields the resonance width of 87.2 keV, which agrees well 
with the experimental value, 86 (6) keV \cite{Kondo15}. 

In addition to the $d_{3/2}$ resonance, we also find a broad $p_{3/2}$ and 
a relatively narrow $f_{7/2}$ resonances with this potential. 
For the $p_{3/2}$ resonance, the resonance energy and the width are 
$E$ = 0.577 MeV and $\Gamma$ = 1.63 MeV, respectively, while they are 
$E$ = 2.44 MeV and $\Gamma$ = 0.21 MeV for the $f_{7/2}$ resonance. 
Notice that, due to the lower centrifugal barrier, the 
$p_{3/2}$ resonance appears at a lower energy with a larger width 
compared to the $d_{3/2}$ and $f_{7/2}$ resonances. 
The existence of the three resonance states 
in $^{25}$O is consistent with 
a prediction reported in Ref. \cite{Lepailleur15} based on 
shell model and Skyrme Hartree-Fock calculations, although 
the resonance widths are not evaluated there (see Table III and 
Fig. 12 in 
Ref. \cite{Lepailleur15}).
The energy and the width of each of these three resonance states 
are summarized in Table I. 
In Sec. III, we will discuss the structure of excited 0$^+$ states 
of $^{26}$O in connection to these single-particle resonance states of 
$^{25}$O. 

\begin{table}[bt]
\caption{Single-particle resonance states of the $^{25}$O nucleus 
obtained with a $n$+$^{24}$O model. The Woods-Saxon potential 
is calibrated using the energy of the $d_{3/2}$ resonance. 
The resonance energy, $E$, and the width, $\Gamma$, are obtained by 
imposing the outgoing wave boundary condition to the radial wave function 
for each angular momentum $j$ and $l$.}
\begin{tabular}{c|cc}
\hline\hline
angular momentum & $E$ (MeV) & $\Gamma$ (MeV)   \\
\hline 
d$_{3/2}$ & 0.749 (input) & 0.0872 \\
p$_{3/2}$ & 0.577 & 1.63 \\
f$_{7/2}$ & 2.44 & 0.21 \\
\hline
expt. d$_{3/2}$ \cite{Kondo15} & 0.749 (10) & 0.088 (6) \\
\hline\hline
\end{tabular}
\end{table}

\subsection{One-particle Green's function and the resonance width}

While we investigated in the previous subsection the resonance 
structure of the $^{25}$O nucleus using the Gamow states with complex 
eigen-energies, the resonance structure can also be studied using the 
Green's function keeping the energy to be real. 
In this approach, the decay energy spectrum is given by 
\begin{equation}
\frac{dP}{dE} = |\langle \Phi_{\rm ref}|\psi_E \rangle|^2, 
\label{dPdE0}
\end{equation}
where $\Phi_{\rm ref}$ is the wave function for a reference state. 
$\psi_E$ is a continuum wave function at $E$ for a Hamiltonian of interest 
and is given by 
\begin{equation}
\psi_E(\vec{r}) = \frac{u_{jl}(r)}{r}{\cal Y}_{jlm}(\hat{\vec{r}}), 
\end{equation}
with 
\begin{equation}
u_{jl}(r)\to \sqrt{\frac{2\mu}{\pi k\hbar^2}}\,
\sin\left(kr-\frac{l}{2}\pi+\delta_{jl}(E)\right)~~~~~(r\to\infty). 
\label{wf-asympt}
\end{equation}
Here, ${\cal Y}_{jlm}(\hat{\vec{r}})$ is the spin-angular wave function, 
$k=\sqrt{2\mu E/\hbar^2}$ is the wave number, and 
$\delta_{jl}(E)$ is the phase shift. 
The normalization factor in Eq. (\ref{wf-asympt}) is chosen such that the wave 
function $\psi_E$ satisfies the normalization condition of 
$\int dE \,|\psi_E\rangle\langle \psi_E| = 1$. 
Equation (\ref{dPdE0}) indicates that the decay energy spectrum $dP/dE$ increases 
when the overlap between the reference state and the continuum state is 
large. Therefore, if one chooses the reference wave function to be well 
confined inside a barrier, the decay energy spectrum shows a peak around 
the resonance energy, at which there is an appreciable component of 
the continuum wave function inside the barrier. 
The reference state is referred to as a doorway state in Ref. \cite{TOF15}. 

The decay spectrum, Eq. (\ref{dPdE0}), can also be expressed in a different 
way using the relation 
\begin{equation}
\lim_{\eta\to0}\frac{1}{x-i\eta}=\frac{1}{x}+i\pi\delta(x). 
\end{equation}
That is, 
\begin{eqnarray}
\frac{dP}{dE} &=& 
\int dE' 
|\langle \Phi_{\rm ref}|\psi_{E'} \rangle|^2\,\delta(E'-E), \\
&=&
\frac{1}{\pi}\,{\rm Im}\int dE' 
|\langle \Phi_{\rm ref}|\psi_{E'} \rangle|^2\,
\frac{1}{E'-E-i\eta},
\label{dPdE2}
\end{eqnarray}
where Im denotes the imaginary part and $\eta$ is 
taken to be an infinitesimal number. 
Notice that 
\begin{equation}
\int dE'\,|\psi_{E'} \rangle\,\frac{1}{E'-E-i\eta}\,\langle\psi_{E'}|
=\frac{1}{\hat{h}-E-i\eta}
\end{equation}
is nothing but the Green's function, $G(E)$. 
Equation (\ref{dPdE2}) can therefore be written also as 
\begin{equation}
\frac{dP}{dE} = \frac{1}{\pi}\,{\rm Im}
\langle \Phi_{\rm ref}|G(E)|\Phi_{\rm ref} \rangle. 
\label{Green}
\end{equation}

\begin{figure} [tb]
\includegraphics[scale=0.5,clip]{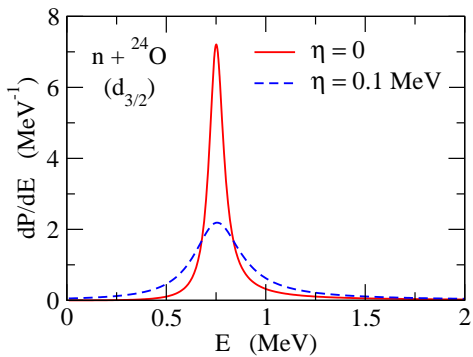}
\caption{(Color online) 
The decay energy spectrum for the $d_{3/2}$ 
configuration for the 
$n$+$^{24}$O system. The solid line is obtained with Eq. (\ref{dPdE0}) 
while the dashed line is obtained using Eq. (\ref{dPdE2}) with a finite 
value of $\eta=0.1$ MeV. 
The reference state $\Phi_{\rm ref}$ 
is taken to be a bound neutron $d_{3/2}$ state in the $^{25}$F nucleus.}
\label{fig:dPdE0}
\end{figure}

The solid line in Fig. \ref{fig:dPdE0} shows the decay energy spectrum 
for the $d_{3/2}$ configuration of the $^{25}$O nucleus. To draw this curve, 
we use the neutron 1$d_{3/2}$ state at $\epsilon_{1d_{3/2}}=-0.811$ MeV 
in the $^{26}$F nucleus for the reference state, $\Phi_{\rm ref}$. 
To this end, we use a similar potential as $V_{nC}$ for the $^{25}$O 
nucleus, but by modifying the strength of the spin-orbit potential 
to be $V_{ls}$=33.50 MeV fm$^2$ taking into account the tensor force 
between the valence proton and neutron \cite{OSHSA10,O05,SBF77,CSFB07,LBB07}. 
As is expected, the decay energy spectrum shows a sharp peak at $E=0.75$ 
MeV. The curve is approximately given by the Breit-Wigner function with 
a natural width of 0.087 MeV, which is consistent with the one obtained 
with the Gamow state method shown in the previous subsection (see Table I). 

\begin{figure} [tb]
\includegraphics[scale=0.5,clip]{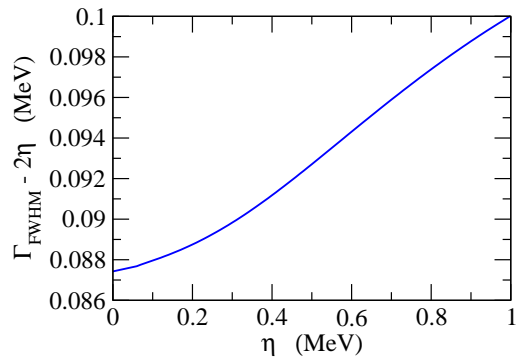}
\caption{(Color online) 
The $d_{3/2}$ resonance width 
for the $n+^{24}$O system 
estimated from the 
decay energy spectrum with a finite value of 
$\eta$ in the Green's function. 
}
\label{fig:FWHM}
\end{figure}

The dashed line in Fig. \ref{fig:dPdE0}, on the other hand, is 
obtained with Eq. (\ref{dPdE2}) by keeping a finite value 
of $\eta$= 0.1 MeV. 
One can see that the peak position remains almost the same as in 
the solid line, but the width increases significantly because of the 
smearing factor $1/(E'-E-i\eta)$ in Eq. (\ref{dPdE2}). 
If one approximates the decay energy spectrum in the limit of $\eta\to$0 
by the Breit-Wigner function with the resonance energy $E_R$ and 
the natural width $\Gamma$, that is, 
\begin{equation}
|\langle \Phi_{\rm ref}|\psi_{E} \rangle|^2
\sim\frac{1}{2\pi}\,\frac{\Gamma}{(E-E_R)^2+\frac{\Gamma^2}{4}},
\end{equation}
Eq. (\ref{dPdE2}) is written as 
\begin{equation}
\frac{dP}{dE} \sim \int dE' 
\,\frac{\frac{\Gamma}{2\pi}}{(E'-E_R)^2+\frac{\Gamma^2}{4}}
\cdot\frac{\frac{\eta}{\pi}}{(E-E')^2+\eta^2}. 
\end{equation}
It is known that a convolution of the Breit-Wigner function with 
another Breit-Wigner function is again a Breit-Wigner function with 
the same resonance energy and the sum of the two resonance widths 
(this can be easily confirmed by performing the Fourier transform 
and then the inverse Fourier transform back to the 
original function). 
This equation thus becomes, 
\begin{equation}
\frac{dP}{dE} \sim 
\frac{\frac{\Gamma+2\eta}{2\pi}}{(E-E_R)^2+\frac{(\Gamma+2\eta)^2}{4}}. 
\label{convolution}
\end{equation}
This implies that the resonance width can numerically be estimated 
with the decay energy spectrum calculated with a finite value of $\eta$ 
as 
\begin{equation}
\Gamma = \Gamma_{\rm FWHM}-2\eta,
\label{FWHM}
\end{equation}
where $\Gamma_{\rm FWHM}$ is the full width at half maximum (FWHM) of the calculated 
spectrum, Eq. (\ref{dPdE2}). 

In order to test this idea, Fig. \ref{fig:FWHM} shows the right hand side 
of Eq. (\ref{FWHM}) as a function of $\eta$. 
One can see that this method yields the resonance width within the accuracy 
of about 1\% at $\eta$=0.1 MeV. In practice, one may extrapolate 
the values for 
different $\eta$ down to $\eta$ = 0 
in order to estimate the resonance width. 
This method is convenient particularly for the three-body problem which 
we shall discuss in the next section. 

\section{Decay energy spectrum for $0^+$ states of the $^{26}$O nucleus}

\subsection{Decay energy spectrum}

Let us now solve a three-body Hamiltonian for the $^{26}$O nucleus based on the $^{24}$O + $n$ + $n$ 
model and discuss the decay dynamics. To this end, we consider the following three-body Hamiltonian, 
\begin{equation}
H=h_{nC}(1)+h_{nC}(2)+v(\vec{r}_1,\vec{r}_2), 
\end{equation}
where $h_{nC}$ is the single-particle Hamiltonian given by Eq. (\ref{Hsp}) and 
the pairing interaction is taken to be a density dependent contact 
interaction as \cite{BE91,EBH97,EB92,VinhMau96,HS05,SH15}, 
\begin{equation}
v(\vec{r}_1,\vec{r}_2)=\delta(\vec{r}_1-\vec{r}_2)
\left(v_0+\frac{v_\rho}{1+\exp[(r_1-R_\rho)/a_\rho]}\right). 
\label{vnn}
\end{equation}
For simplicity, we have neglected the two-body part of the recoil kinetic energy 
of the core nucleus, as in our previous works \cite{HS14,HS14-2}. 
In the density dependent pairing interaction, Eq. (\ref{vnn}), the strength 
of the density independent part is given as \cite{EBH97}, 
\begin{equation}
v_0=2\pi^2\frac{\hbar^2}{m_N}\,\frac{2a_{nn}}{\pi-2k_Ca_{nn}}, 
\end{equation}
where $a_{nn}$ is the scattering length for the $nn$ scattering and $k_C$ is 
related to the cut-off energy, $E_{\rm cut}$, as $k_C=\sqrt{m_NE_{\rm cut}/\hbar^2}$. 
Following Ref. \cite{EBH97}, we take $a_{nn}=-15$ fm. With $E_{\rm cut}$ = 30 MeV, 
this leads to $v_0 = -857.2$ MeV fm$^3$. 
For the parameters for the density dependent part in Eq. (\ref{vnn}), we determine them 
so as to reproduce the ground state energy of $^{26}$O, $E$ = 18 keV \cite{Kondo15}. 
The values of the parameters which we employ are $R_\rho$ = 1.34 $\times A_c^{1/3}$ fm, 
$a_\rho$ = 0.72 fm, and $v_\rho$ = 928.95 MeV fm$^3$. 

As in the previous section, the decay energy spectrum is obtained with the Green's method, 
Eq. (\ref{Green}), with some three-body wave function for $\Phi_{\rm ref}$ and the two-particle 
Green's function given by $G(E)=1/(H-E-i\eta)$. 
We evaluate the correlated Green's function, $G(E)$, using the relation \cite{EB92},
\begin{equation}
G(E) = G_0(E)-G_0(E)v(1+G_0(E)v)^{-1}G_0(E),
\label{Green2}
\end{equation}
where the uncorrelated two-particle Green's function, $G_0(E)$, is given by 
\begin{eqnarray}
G_0(E)&=&\frac{1}{h_{nC}(1)+h_{nC}(2)-E-i\eta}, \\ 
&=&\sum_{j_1,l_1}\sum_{j_2,l_2}\int de_1de_2\,\frac{|[\psi_1\psi_2]^{(0^+)}\rangle\langle 
[\psi_1\psi_2]^{(0^+)}|}{e_1+e_2-E-i\eta}. \nonumber \\
\label{Green0}
\end{eqnarray}
Since the interaction $v$ is zero-ranged, the inversion of the operator $(1+G_0(E)v)$ in 
Eq. (\ref{Green2}) is best performed in the coordinate space on a finite radial 
grid \cite{BE91,EB92}. 
With Eqs. (\ref{Green2}) and (\ref{Green0}), 
the uncorrelated spectrum is then given by 
\begin{eqnarray}
\frac{dP_0}{dE}&=&
\frac{1}{\pi}\,{\rm Im}
\langle \Phi_{\rm ref}|G_0(E)|\Phi_{\rm ref} \rangle, \\
&=&
\frac{1}{\pi}\,{\rm Im}
\sum_{j_1,l_1}\sum_{j_2,l_2}\int de_1de_2\,
\frac{|\langle\Phi_{\rm ref}|[\psi_1\psi_2]^{(0^+)}\rangle|^2}{e_1+e_2-E-i\eta}, \nonumber \\
\end{eqnarray}
while the correlated spectrum is evaluated as \cite{EB92},
\begin{eqnarray}
\frac{dP}{dE}&=&\frac{dP_0}{dE} -\frac{1}{\pi}\,{\rm Im}\int d\vec{r}d\vec{r}'
\tilde{G}_D(\vec{r})v(\vec{r}) \nonumber \\
&&\times (1+G_0(E)v)^{-1}_{\vec{r}\vec{r}'}G_D(\vec{r}'),
\end{eqnarray}
with 
\begin{eqnarray}
G_D(\vec{r})&=&
\sum_{j_1,l_1}\sum_{j_2,l_2}\int de_1de_2\, \nonumber \\
&&\times \frac{\langle\vec{r}\vec{r}|[\psi_1\psi_2]^{(0^+)}\rangle\langle 
[\psi_1\psi_2]^{(0^+)}|\Phi_{\rm ref}\rangle}{e_1+e_2-E-i\eta},
\end{eqnarray}
and
\begin{eqnarray}
\tilde{G}_D(\vec{r})&=&
\sum_{j_1,l_1}\sum_{j_2,l_2}\int de_1de_2\, \nonumber \\
&&\times \frac{\langle\Phi_{\rm ref}|[\psi_1\psi_2]^{(0^+)}\rangle\langle 
[\psi_1\psi_2]^{(0^+)}|\vec{r}\vec{r}\rangle}{e_1+e_2-E-i\eta}.
\end{eqnarray}
Notice that $\tilde{G}_D(\vec{r})$ is not identical to  
$G^\dagger_D(\vec{r})$. 

\begin{figure} [tb]
\includegraphics[scale=0.5,clip]{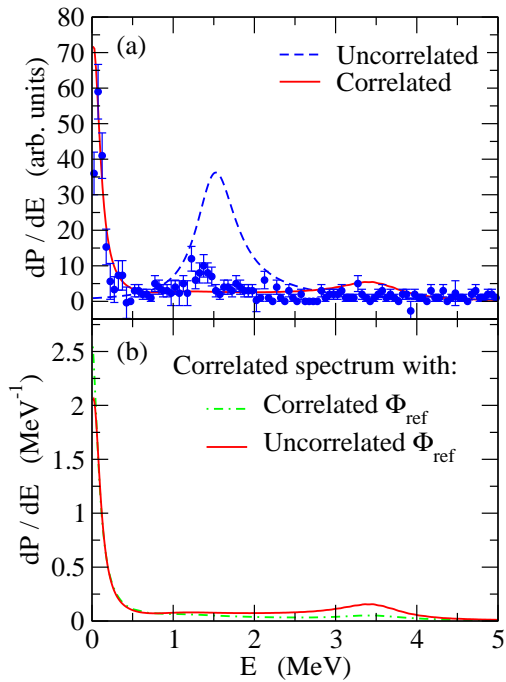}
\caption{(Color online) 
Upper panel: the uncorrelated (the dashed line) and the correlated (the 
solid line) decay energy spectra for the $^{26}$O nucleus. 
The uncorrelated two-neutron state in $^{27}$F with 
the $|[1d_{3/2}1d_{3/2}]^{(0^+)}\rangle$ configuration is employed 
for the reference state, $\Phi_{\rm ref}$. For a presentation purpose, 
a finite value of $\eta$ = 0.1 MeV is used. 
The experimental data 
are taken from Ref. \cite{Kondo15}. 
Lower panel: the dependence of the correlated decay spectrum 
on the choice of the reference state. The solid line is the same as that in 
the upper panel and is obtained with the uncorrelated two-particle 
wave function of $^{27}$F. The dot-dashed line, on the other hand, is obtained with 
the correlated two-particle wave function of $^{27}$F based on the $^{25}$F + $n$ + $n$ 
three-body model. 
}
\label{fig:decayspectrum}
\end{figure}

The upper panel of Fig.  \ref{fig:decayspectrum} shows the 
uncorrelated (the dashed line) and the correlated (the solid line) 
decay spectra for $^{26}$O. 
To this end, we use 
the uncorrelated two-neutron state of $^{27}$F with the  
$|[1d_{3/2}1d_{3/2}]^{(0^+)}\rangle$ configuration 
for the reference state $\Phi_{\rm ref}$, since the $^{26}$O nucleus 
was produced in the single proton-knockout reaction from a secondary 
$^{27}$F beam in the experiments of Refs. \cite{LDK12,CSA13,Kondo15}. 
For a presentation purpose, we keep a finite value of $\eta$ =0.1 MeV 
in Eq. (\ref{Green0}) in order to evaluate the uncorrelated 
Green's function. 
In fact, numerically it is much easier to evaluate the correlated 
Green's function with a finite value of $\eta$, 
especially when a natural width is small, 
since in any case one must discretize the energies in the quadratures 
in Eq. (\ref{Green0}). 
In the figure, the main feature of the decay energy spectra 
is the same as 
that in Ref. \cite{HS14}. That is, a peak appears at twice the 
single-particle resonance energy, 1.498 MeV, in the uncorrelated spectrum, 
which is largely shifted towards the threshold energy in the correlated 
spectrum due to the pairing correlation. 

The lower panel of Fig. \ref{fig:decayspectrum} shows the dependence of 
the result on the choice of the reference state. The solid line is the 
same as that in the upper panel, while the dot-dashed line is obtained 
with the correlated three-body wave function for the $^{27}$F nucleus 
obtained with the $^{25}$F + $n$ + $n$ model using the same pairing 
interaction as in Eq. (\ref{vnn}). 
One can see that qualitatively 
the spectrum does not depend much on the choice of 
the reference state. In particular, the sharp low-energy peak 
is produced in both the calculations, although the height is 
somewhat lower in the calculation with the uncorrelated reference state. 
This is because the ground state is mainly 
composed of the 
$|[1d_{3/2}1d_{3/2}]^{(0^+)}\rangle$ configuration 
both in the $^{26}$O and in the $^{27}$F nuclei (see the next subsection). 

This finding implies that the decay energy spectrum does not depend much on how 
the $^{26}$O nucleus is produced. 
Intuitively, for a narrow resonance with a long lifetime, 
a nucleus looses its memory of how it was produced, 
before it decays under the barrier
\cite{GECZ12}. 
In this situation, the decay dynamics predominantly 
reflects properties of the continuum wave function of the final three-body 
system, $\Psi_E$. 
In the terminology of nuclear reaction, the decay involves only the final 
state interactions. 
This is in fact a good point of narrow resonances, since one 
does not have to bother with the reaction dynamics and/or the structure of 
the initial nucleus in analyzing 
observables in the decay process. We will discuss an example of this 
point 
in Sec. IV, that is, the angular correlation of the two emitted neutrons, 
which reflects the dineutron correlation in 
the continuum state of $^{26}$O, rather than the properties of the initial 
state of $^{27}$F. 

In the following calculations, for simplicity, we shall use the uncorrelated
wave function for the reference state $\Phi_{\rm ref}$. 

\subsection{Di-neutron correlation in the ground state}

Since the resonance width is 
considerably small for the ground state of $^{26}$O \cite{KLD13,GMZ13}, 
one would expect that a bound state approximation provides a reasonable 
result in discussing the ground state properties of the $^{26}$O nucleus. 
Let us therefore obtain the ground state wave function by putting the 
$^{26}$O nucleus in a large box as in Ref. \cite{HS05}. 
To this end, we use the box size of $R_{\rm box}$ = 40 fm. 
Since the continuum states are treated approximately in this calculation, 
we slightly readjust the $v_\rho$ parameter 
in the pairing interaction, 
Eq. (\ref{vnn}), so that the resultant ground state energy is 18 keV. 

\begin{table}[bt]
\caption{Ground state properties of the $^{26}$O nucleus obtained with 
the bound state approximation with a box size of $R_{\rm box}$ = 40 fm. 
$\langle r_{nn}^2\rangle$ and $\langle r_{c-2n}^2\rangle$ 
are the mean-square neutron-neutron distance and the core-dineutron 
distance, respectively. The fractions of the main components 
and the spin-singlet component are also shown. }
\begin{tabular}{cccccccc}
\hline\hline
$\langle r_{nn}^2\rangle$ & $\langle r_{c-2n}^2\rangle$ & $[d_{3/2}]^2$ &
$[f_{7/2}]^2$ & $[p_{3/2}]^2$ & $[p_{1/2}]^2$ & $[g_{9/2}]^2$ 
& $S=0$ \\
(fm$^2$) & (fm$^2$) & (\%) & (\%) & (\%) & (\%)  & (\%) & (\%) \\
\hline 
40.2 & 18.5 & 66.1 & 18.3 & 10.5 & 1.50 & 1.15 & 54.8 \\
\hline\hline
\end{tabular}
\end{table}

\begin{figure} [tb]
\includegraphics[scale=0.5,clip]{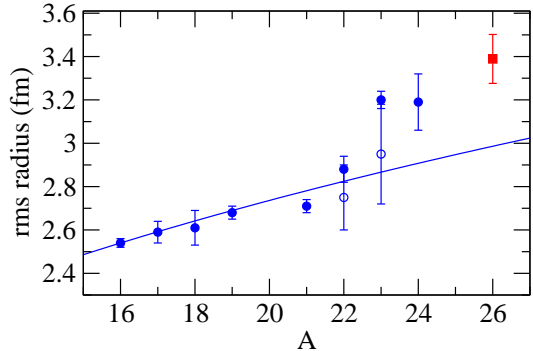}
\caption{(Color online) 
The root-mean-square (rms) radii of the oxygen isotopes as a function of the 
mass number $A$. 
The filled and open circles are the experimental data taken from Refs. 
\cite{Ozawa01} and \cite{Kanungo11}, respectively. 
The filled square is the radius of the $^{26}$O estimated with the three-body 
model calculation with the empirical rms radius of $^{24}$O. The solid 
line shows a function $\sqrt{\langle r^2\rangle} = r_0 A^{1/3}$ with 
$r_0$ = 1.0 fm, in which the value of $r_0$ is determined with the radius of $^{16}$O.
}
\label{fig:Oradii}
\end{figure}

Table II summarizes the properties of the ground state wave function thus obtained. 
One can see that the ground state wave function mainly consists of the 
$(d_{3/2})^2$ configuration, but there is also an appreciable admixture of other 
components, such as the $(f_{7/2})^2$ and the $(p_{3/2})^2$ 
components (see also Ref.\cite{Lepailleur15}). 
From the calculated root-mean-square (rms) inter-neutron distance, 
$\sqrt{\langle r_{nn}^2\rangle}$, and the core-dineutron distance, 
$\sqrt{\langle r_{c-2n}^2\rangle}$, one can estimate the radius of the 
$^{26}$O as \cite{VinhMau96}, 
\begin{equation}
\langle r^2\rangle_{^{26}{\rm O}} = 
\frac{24}{26}\,\langle r^2\rangle_{^{24}{\rm O}} +\delta \langle r^2\rangle, 
\end{equation}
with 
\begin{equation}
\delta \langle r^2\rangle=\frac{1}{26}\,\left(\frac{2\times 24}{26} 
\langle r_{c-2n}^2\rangle + \frac{1}{2}\langle r_{nn}^2\rangle \right). 
\end{equation}
Using the empirical rms radius of $^{24}$O, 
$\sqrt{\langle r^2\rangle}_{^{24}{\rm O}} = 3.19 \pm$ 0.13 fm \cite{Ozawa01}, 
obtained with the interaction cross section measurement, we estimate 
the rms radius of $^{26}$O to be 
$\sqrt{\langle r^2\rangle}_{^{26}{\rm O}} = 3.39 \pm$ 0.11 fm. 
The radii of the oxygen isotopes are shown in Fig. \ref{fig:Oradii}. 
The filled and open circles are the experimental radii taken from Refs. 
\cite{Ozawa01} and \cite{Kanungo11}, respectively, while the filled square 
is the calculated radius for $^{26}$O. 
One can see that the radius of the $^{26}$O nucleus is significantly 
larger than the empirical law
of $r\propto A^{1/3}$, as is shown 
by the solid line in the figure. This may suggest a halo structure of the 
$^{26}$O nucleus.

\begin{figure} [tb]
\includegraphics[scale=0.5,clip]{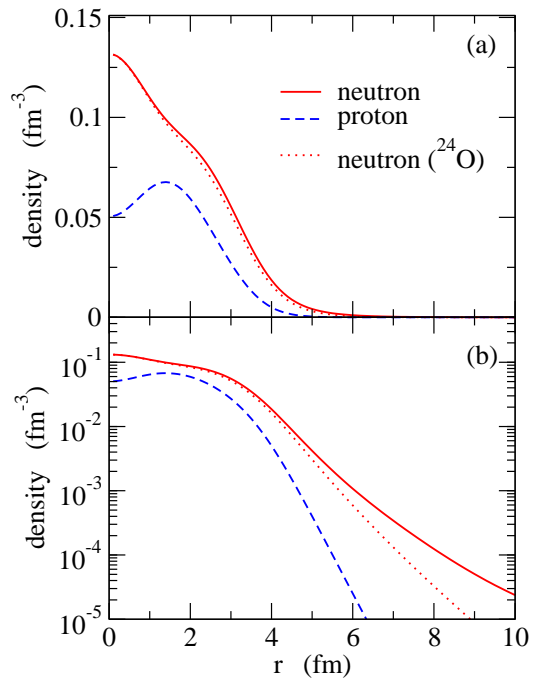}
\caption{(Color online) 
The density distribution of the $^{26}$O nucleus in the linear (the upper panel) and 
in the logarithmic (the lower panel) scales, respectively. 
The solid and the dashed lines denote the neutron and the proton densities, respectively, while 
the dotted line shows the neutron density of the core nucleus, $^{24}$O. 
The density distribution for the valence neutrons is obtained with the three-body model while the 
density of the core nucleus is constructed using the Skyrme-Hartree-Fock calculation.
}
\label{fig:density1}
\end{figure}

In order to see whether $^{26}$O has a halo structure, Fig. \ref{fig:density1} shows the density distribution 
of the $^{26}$O nucleus. To this end, we construct the density of the core nucleus, $^{24}$O, with the Skyrme-Hartree-Fock calculation with SLy4 interaction \cite{SLy4}. We do not take into account the pairing correlation 
in $^{24}$O in order to be consistent with the three-body model for $^{26}$O. This calculation yields 3.05 fm for the 
matter radius of $^{24}$O, which agrees with the experimental value within the error bar. 
The upper and the lower panels in the figure 
show the density distribution in linear and logarithmic 
scales, respectively. 
One can see that the neutron distribution (the solid line) is more extended than 
the proton density (the dashed line) and than 
the neutron density in the core nucleus (the dotted line). 

\begin{figure} [tb]
\includegraphics[scale=1.3,clip]{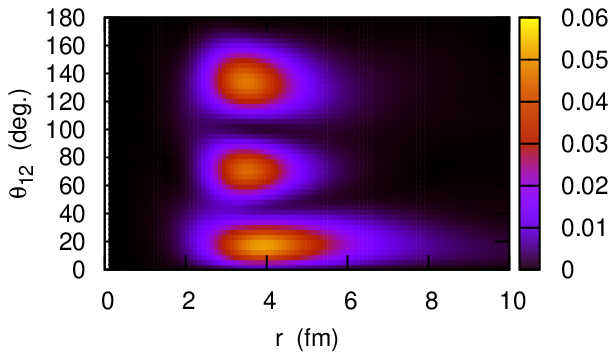}
\caption{(Color online) 
The two-particle density for the $^{26}$O nucleus as a function of 
$r_1=r_2=r$ and the angle between the valence neutrons, $\theta_{12}$. 
It is weighted with a factor of $8\pi^2r^4\sin\theta_{12}$. 
}
\label{fig:density2}
\end{figure}

\begin{figure} [tb]
\includegraphics[scale=1.3,clip]{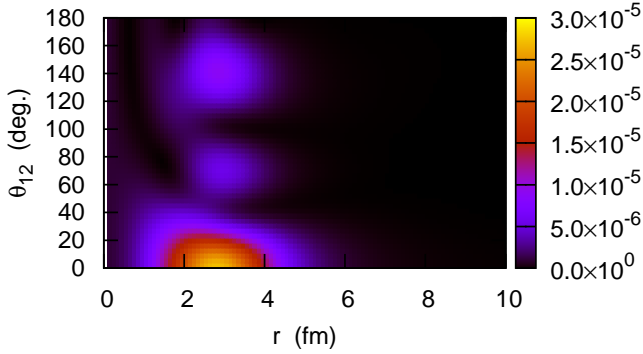}
\caption{(Color online) 
Same as Fig. \ref{fig:density2}, but without the weight factor of 
$8\pi^2r^4\sin\theta_{12}$. 
}
\label{fig:density2-2}
\end{figure}

\begin{figure} [tb]
\includegraphics[scale=0.4,clip]{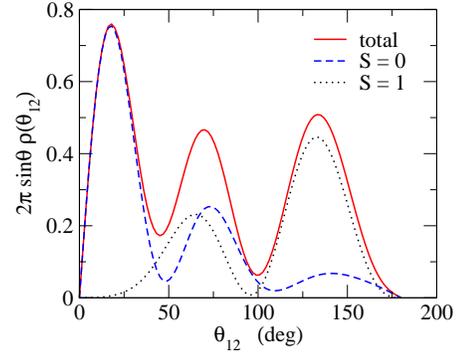}
\caption{(Color online) 
The angular density (weighted with a factor $2\pi\sin\theta_{12}$) 
for the ground state of the $^{26}$O nucleus 
as a function of the angle between the two valence neutrons, $\theta_{12}$. 
The solid line is for the total density, while the dashed and the dotted 
lines are for the spin-singlet $(S=0)$ and the spin-triplet $(S=1)$ components, respectively. }
\label{fig:angdensity}
\end{figure}

Figs. \ref{fig:density2} and \ref{fig:density2-2} 
show the two-particle density, $\rho(r_1,r_2,\theta_{12})$, 
of the $^{26}$O nucleus with and without 
the weight factor of 
$8\pi^2r^4\sin\theta_{12}$, respectively. These are plotted as a function 
of $r_1=r_2=r$ and the angle between the valence neutrons, $\theta_{12}$. 
One can see, especially in Fig. \ref{fig:density2-2}, that the two-particle 
density is well concentrated in the small $\theta_{12}$, which is 
a clear manifestation of the dineutron correlation \cite{HS05}. 
Notice that, with the Woods-Saxon potential 
which we employ, the position of the centrifugal barrier is at $r$ = 6.08 fm for $d_{3/2}$. 
The dineutron correlation is thus present inside the barrier, hence before the two-neutron decay.  
The angular density defined by \cite{HS05}, 
\begin{equation}
\rho(\theta_{12}) = 
4\pi\int^\infty_0 r_1^2dr_1\int^\infty_0 r_2^2dr_2\,
\rho(r_1,r_2,\theta_{12}),
\end{equation}
is plotted in Fig. \ref{fig:angdensity} as a function of $\theta_{12}$. 
The figure also shows a decomposition into the spin-singlet ($S$ = 0) 
and and the spin-triplet ($S$ = 1) components \cite{BE91}. 
One can see that the dineutron component in the small $\theta_{12}$ region 
consists predominantly of the $S$ = 0 component, while the peak in the 
large $\theta_{12}$ region mainly consists of the $S$ = 1 component. 
This has some similarity to the $^{11}$Li nucleus shown in Ref. \cite{HS05}, 
although the middle peak is absent in $^{11}$Li due to the dominance of the $(p_{1/2})^2$  configuration 
instead of the $(d_{3/2})^2$ configuration. 

In connection to the angular correlations of the two emitted neutrons, which we will  
discuss in Sec. V, it is interesting here 
to discuss the ground state density 
in the momentum space. 
Suppose that one expresses the ground state wave function in 
the coordinate space 
as \cite{BE91,EBH97,HS05}
\begin{equation}
\Psi(\vec{r}_1,\vec{r}_2)=\sum_{n,n'}\sum_{l,j}\alpha_{nn'lj}
\Psi_{nn'lj}(\vec{r}_1,\vec{r}_2),
\label{wf2-r}
\end{equation}
with 
\begin{equation}
\Psi_{nn'lj}(\vec{r}_1,\vec{r}_2) 
=\sum_m\langle jmj-m|00\rangle \psi_{njlm}(\vec{r}_1)\psi_{n'jl-m}(\vec{r}_2), 
\end{equation}
where $n$ is the radial node and the single-particle wave function is 
given by 
\begin{equation}
\psi_{njlm}(\vec{r})=\phi_{njl}(r){\cal Y}_{jlm}(\hat{\vec{r}}). 
\end{equation}
Noticing that the Fourier transform of 
$\psi_{njlm}(\vec{r})$ is given by 
\begin{eqnarray}
\tilde{\psi}_{njlm}(\vec{k})&=&
\int d\vec{r}\,e^{i\vec{k}\cdot\vec{r}}
\psi_{njlm}(\vec{r})=i^l\tilde{\phi}_{njl}(k)
{\cal Y}_{jlm}(\hat{\vec{k}}), \nonumber \\
\\
&\equiv& i^l  \tilde{\psi}'_{njlm}(\vec{k})
\end{eqnarray}
with 
\begin{equation}
\tilde{\phi}_{njl}(k)
=4\pi \int r^2dr \,j_l(kr)\phi_{njl}(r),
\end{equation}
the Fourier transform of $\Psi(\vec{r}_1,\vec{r}_2)$ reads 
\begin{equation}
\tilde{\Psi}(\vec{k}_1,\vec{k}_2)=\sum_{n,n'}\sum_{l,j}(-1)^l\alpha_{nn'lj}
\tilde{\Psi}_{nn'lj}(\vec{k}_1,\vec{k}_2),
\label{wf2-p}
\end{equation}
with 
\begin{equation}
\tilde{\Psi}_{nn'lj}(\vec{k}_1,\vec{k}_2) 
=\sum_m\langle jmj-m|00\rangle \tilde{\psi}'_{njlm}(\vec{k}_1)
\tilde{\psi}'_{n'jl-m}(\vec{k}_2). 
\end{equation}
By comparing Eq. (\ref{wf2-r}) with Eq. (\ref{wf2-p}), one can notice 
that the role of odd-partial waves is opposite between the coordinate 
space and the momentum space. Notice that the dineutron correlation is 
caused by the coherent superposition between even- and odd- partial 
waves \cite{PSS07,Catara84,HV11}. 
If the weight factors $\alpha_{nn'lj}$ in Eq. (\ref{wf2-r}) are such that 
the wave function for $\vec{r}_1=\vec{r}_2$ is enhanced and that for 
$\vec{r}_1=-\vec{r}_2$ is suppressed, as in the dineutron correlation, 
the wave function in the momentum space therefore 
shows an enhancement for $\vec{k}_1=-\vec{k}_2$ and a suppression for 
$\vec{k}_1=\vec{k}_2$. 
This fact can also be understood in terms of the uncertainty relation 
between the space and the momentum. 

\begin{figure} [tb]
\includegraphics[scale=1.3,clip]{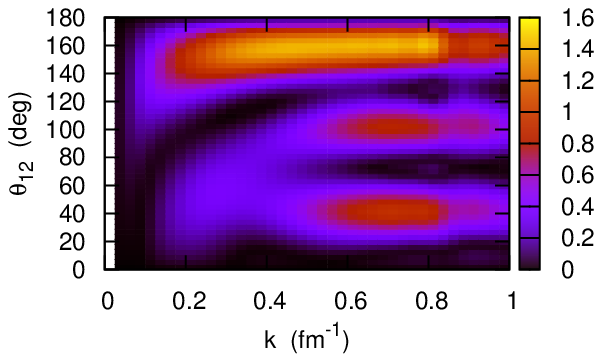}
\caption{(Color online) 
The two-particle density in the momentum space 
as a function of 
$k_1=k_2=k$ and the angle between the valence neutrons, $\theta_{12}$. 
It is weighted with a factor of $8\pi^2k^4\sin\theta_{12}$. 
}
\label{fig:density2-p}
\end{figure}

The two-particle density in the momentum space, 
constructed with $\tilde{\Psi}(\vec{k}_1,\vec{k}_2)$, is shown in 
Fig. \ref{fig:density2-p} as a function of 
$k_1=k_2=k$ and $\theta_{12}$. 
One can clearly see that the two-particle density is indeed enhanced 
in the large $\theta_{12}$ region in the momentum space, reflecting the 
dineutron correlation shown in Fig. \ref{fig:density2} in the 
coordinate space. 

\subsection{Excited 0$^+$ states}

\begin{figure} [tb]
\includegraphics[scale=0.5,clip]{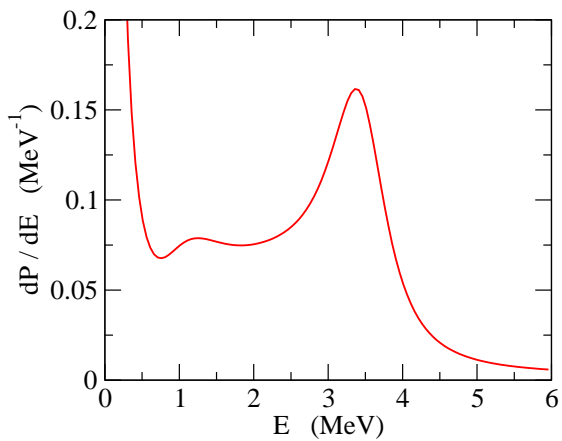}
\caption{(Color online) 
Same as Fig. \ref{fig:decayspectrum}, but plotted in a magnified 
way for the region of $E\geq$ 0.5 MeV. 
}
\label{fig:decayspectrum2}
\end{figure}

The decay energy spectrum shown in Fig. \ref{fig:decayspectrum} shows 
that the ground energy peak 
does not vanish quickly and there is appreciable strength 
between 
$1 \leq E \leq 3 $ MeV, indicating 
the presence of a few unresolved excited 0$^+$ resonances in this energy region. 
In addition, one can see 
a peak at 3.38 MeV. In order to better understand these structures, 
Fig. \ref{fig:decayspectrum2} shows 
the decay energy spectrum in a magnified way.  
One can see at least two peaks, the lower one at 1.22 MeV and the higher one at 3.38 MeV. 
For the higher energy peak, the width is estimated to be 0.737 MeV using the method 
discussed in Sec. II. On the other hand, for the lower energy peak, unfortunately the width cannot be 
estimated, because the 
resonance peak overlaps with other peaks and the full-width-at-half 
maximum cannot be defined. 

In order to understand the structure of these peaks, we evaluate the probabilities of angular 
momentum components in the wave functions. 
Since the correlated Green's function is expressed as 
\begin{equation}
G(E)=\sum_{\widetilde{2p}}\frac{|\widetilde{2p}\rangle\langle\widetilde{2p}|}{E_{2p}-E-i\eta},
\end{equation}
where $|\widetilde{2p}\rangle$ is the correlated two-particle wave function at $E=E_{2p}$, 
the amplitude of the correlated wave function on an uncorrelated basis, $|2p\rangle$, can be 
extracted as 
\begin{equation}
\langle \Phi_{\rm ref}|\widetilde{2p}\rangle \langle\widetilde{2p}|2p\rangle 
={\rm Im} \langle \Phi_{\rm ref}|G(E_{2p})|2p\rangle.
\label{amplitude}
\end{equation}
Notice that many continuum states are degenerate in energy at a given value of $E_{2p}$ 
(this can be easily understood in the uncorrelated limit, in which there are many combinations 
of $(e_1,e_2,j,l)$ leading to the 0$^+$ configuration with the same energy $E=e_1+e_2$). 
By taking the overlap 
$\langle \Phi_{\rm ref}|\widetilde{2p}\rangle$ in Eq. (\ref{amplitude}), 
only those correlated wave functions which have an appreciable overlap with the reference wave 
function, thus the resonance wave functions, contribute on the left hand side. 
Because of this property, when one extracts the amplitudes of resonance wave functions, 
we find that 
this method is more convenient than a similar method presented in 
Ref. \cite{BE91}, which uses the relation 
\begin{equation}
\langle \Phi_{\rm ref}|\delta(\vec{r}_1-\vec{r}_2)|\widetilde{2p}\rangle 
\langle\widetilde{2p}|2p\rangle 
={\rm Im} \langle \Phi_{\rm ref}|\delta(\vec{r}_1-\vec{r}_2)G(E_{2p})|2p\rangle. 
\end{equation}
In practice, we use the uncorrelated basis $|2p\rangle$ obtained with the box discretization 
method (see the previous subsection) with a relatively small box size 
in order to eliminate the contribution outside the centrifugal barrier, 
and evaluate the amplitudes according to, 
\begin{equation}
\langle\widetilde{2p}|2p\rangle 
=\frac{{\rm Im} \langle \Phi_{\rm ref}|G(E_{2p})|2p\rangle}
{\sum_{2p}|{\rm Im} \langle \Phi_{\rm ref}|G(E_{2p})|2p\rangle|^2},
\end{equation}
even though the true continuum wave function is not square integrable. 

\begin{table}[bt]
\caption{
Properties of the ground and excited $0^+$ states in $^{26}$O 
corresponding to the three peaks in the decay energy spectrum. 
For the peak at $E=3.38$ MeV, the resonance width is also evaluated 
using the method presented in Sec. II-B. }
\begin{tabular}{ccccc}
\hline\hline
$E$ & $\Gamma$ & ($p_{3/2})^2$ & ($d_{3/2})^2$ & ($f_{7/2})^2$ \\
(MeV) & (MeV) & (\%) & (\%) & (\%) \\
\hline 
0.018 & - & 10.5 & 66.1 & 18.3 \\
1.22 & - & 60.3 & 26.8 & 2.02 \\
3.38 & 0.737 & 10.4 & 24.9 & 62.1 \\
\hline\hline
\end{tabular}
\end{table}

The results obtained with $R_{\rm box}$ = 15 fm are summarized in Table III. 
The table also shows the result for the ground state obtained in the previous 
subsection with the box discretization method. 
One can see that these three resonance peaks nicely reflect the single-particle 
resonance states of the two-body subsystem, $n$+$^{24}$O, shown in Table I. 
That is, the first (i.e., the ground state), the second, and the third peaks in the 
decay energy spectrum mainly consist of the $(d_{3/2})^2$, $(p_{3/2})^2$,  and 
$(f_{7/2})^2$ configurations, respectively, as is expected from the single-particle 
resonances. 

In Ref. \cite{GZ15}, Grigorenko and Zhukov obtained two excited 0$^+$ states at 
1.7 and 2.6 MeV, in addition to the ground state resonance at 0.01 MeV. 
All these resonance states show the dominance of the $(d_{3/2})^2$ configurations, 
and thus 
their excited 0$^+$ states are apparently different from the resonance peaks found 
in the present calculation. As we have mentioned, 
unresolved resonances may exist around 2 MeV in the decay spectrum shown 
in Fig. \ref{fig:decayspectrum2}, 
and the resonance states found by Grigorenko and Zhukov might correspond to some of these. 
As for the $(f_{7/2})^2$ resonance at 
3.38 MeV, 
it is not clear whether they have included the $l$ = 3 configuration 
in the $n$-$^{24}$O channel,  
since they have mentioned the $n$-$^{24}$O potential only for the $s$, $p$, 
and $d$ waves \cite{GZ15}. The correspondence 
is thus not certain at this moment. 

\section{The $2^+$ state}

\begin{figure} [tb]
\includegraphics[scale=0.5,clip]{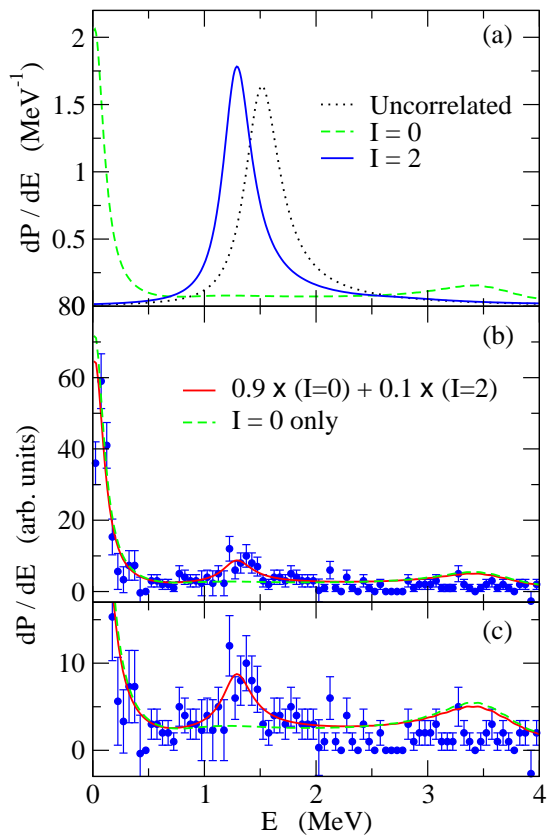}
\caption{(Color online) 
Top panel: 
The decay energy spectrum of the $^{26}$O nucleus. 
The dashed and the solid lines are for the 0$^+$ and 2$^+$ states, respectively, while the dotted line shows 
the uncorrelated spectrum. 
Middle panel: The decay energy spectrum obtained by mixing the 0$^+$ and the 2$^+$ components, as is indicated 
in the figure. 
The experimental data are taken from Ref. \cite{Kondo15}. 
Bottom panel: same as the middle panel, but with a larger scale. 
}
\label{fig:decayspectrum-2+}
\end{figure}

Let us next discuss the first 2$^+$ state in $^{26}$O. One of the most important findings in the 
recent experiment reported in Ref. \cite{Kondo15} is a finding of a clear second peak at 
$E$ = 1.28$^{+0.11}_{-0.08}$ MeV, which is likely attributed to the 2$^+$ state. 
A signal of this peak was weak in the earlier experiments, mainly because the statistics were not 
sufficient. In Ref. \cite{HS14-2}, we have investigated the 2$^+$ state in the $^{26}$O nucleus using the 
three-body model. That is, the energy spectrum for this state can still be obtained with the Green's function 
approach, by using a 2$^+$ state for the reference state, $\Phi_{\rm ref}$, as well as 
in the unperturbed Green's function, Eq. (\ref{Green0}) \cite{HS14-2}. Here we repeat the same calculation, but 
with the revised set of input parameters. 
The results for the decay energy spectrum is shown in Fig. \ref{fig:decayspectrum-2+}. 
Again, the main feature remains the same as in Ref. \cite{HS14-2}. 
That is, due to the pairing interaction between the valence neutrons, the energy of the 
2$^+$ state is slightly shifted towards lower energies from the unperturbed energy, whereas the 
energy shift is much larger for the 0$^+$ state due to the larger overlap between the wave functions of 
the two neutrons. The 2$^+$ peak appears at 1.282 MeV, which agrees perfectly with the experimental 
data. 
The middle and the bottom panels of Fig. \ref{fig:decayspectrum-2+} show 
the energy spectrum obtained by mixing 
the 2$^+$ component by 10\% to the 0$^+$ component. 
As has already been shown in Ref. \cite{HS14-2}, the 
experimental data are better 
reproduced by mixing the 2$^+$ component.

\begin{table}[bt]
\caption{
Comparison of the energies of 
the 3/2$^+$ state of $^{25}$O and 
the 2$^+$ state of $^{26}$O obtained with several methods. 
These values, given in units of MeV, are measured from the thresholds.}
\begin{tabular}{cccc}
\hline\hline
method & $^{25}$O (3/2$^+$) & $^{26}$O (2$^+$) & Reference \\
\hline
Shell model (USDA) & 1.301 & 2.4 & \cite{CSA13,BR06} \\
Shell model (USDB) & 1.303 & 2.45 & \cite{CSA13,BR06} \\
chiral NN + 3N & 0.742 & 1.64 & \cite{CSA13} \\
continuum shell model & 1.002 & 1.87 & \cite{VZ06} \\
continuum-coupled shell model & 0.86 & 1.66 & \cite{TOF15} \\
3-body model  & ? & 1.6 & \cite{GZ15} \\ 
3-body model  & 0.749 (input) & 1.282 & this work \\
\hline 
Experiment & 0.749 (10) & 1.28$^{+0.11}_{-0.08}$ & \cite{Kondo15} \\
\hline\hline
\end{tabular}
\end{table}

While we achieve an excellent agreement with the experimental data for the energy of the 2$^+$ state, 
it is striking to notice that most of theoretical calculations performed so far overestimate the energy. 
For instance, 
an ab-initio calculation with chiral $NN$ and 3$N$ interactions predicts 
$E_{2^+}$ to be 1.6 MeV above the ground state \cite{CSA13} (see also 
Ref. \cite{HBCLR13}). 
Shell model calculations with the USDA and USDB interactions \cite{BR06} 
yield the excitation energy of 
1.9 and 2.1 MeV, respectively \cite{CSA13}, 
whereas the continuum shell model 
calculations predict the 2$^+$ energy above the threshold 
to be 1.87 
MeV \cite{VZ06} 
and 1.66 MeV \cite{TOF15}. 
The recent three-body model calculation by Grigorenko and Zhukov shows the energy to be 1.6 MeV \cite{GZ15}. 
We summarize these results in Table IV together with the energy of the 3/2$^+$ state in $^{25}$O 
for each calculation. 
It is not clear what causes these overestimates of the 2$^+$ energy, 
but the 2$^+$ state should certainly appear at an energy slightly lower than the unperturbed state, as long as 
the three-body structure is reasonable. In this sense, 
the ab-initio calculation with chiral $NN$ and 3$N$ interactions 
shows the opposite trend, and the shell model calculations, except for the continuum 
shell model calculations of Refs. \cite{VZ06,TOF15}, 
seem to overestimate the correlation 
(unfortunately, we cannot judge this for the recent 
three-body model calculation of Grigorenko and Zhukov, because they do not discuss the energy of the $^{25}$O 
nucleus and also because the exact form of the spin-orbit form which they employ is not clear). 

\begin{table}[bt]
\caption{
Properties of the 2$^+$ state obtained with the present three-body model. The experimental 
value of the energy is $E_{\rm exp}$ = 1.28$^{+0.11}_{-0.08}$ MeV \cite{Kondo15}. 
}
\begin{tabular}{ccccc}
\hline\hline
$E$  & $\Gamma$ & $(d_{3/2})^2$ & $(p_{1/3}f_{7/2})$ & $(f_{7/2})^2$ \\
(MeV) & (MeV) & (\%) & (\%) & (\%) \\
\hline 
1.282 & 0.121 & 94.6 & 2.06 & 1.05 \\
\hline\hline
\end{tabular}
\end{table}

In addition to the energy of the 2$^+$ state, in this paper we also evaluate the width and the angular momentum 
components using the methods presented in the previous sections. The results are 
$\Gamma$ = 0.121 MeV, and 94.6 \% for the $(d_{3/2})^2$ configuration, 2.06 \% for the $(p_{1/3}f_{7/2})$ configuration, 
and 1.05 \% for the $(f_{7/2})^2$ configuration. These values are summarized in Table V. The 2$^+$ state is 
predominantly consists of the  
$(d_{3/2})^2$ configuration, which supports our three-body model argument of the energy of 
the 2$^+$ state \cite{HS14-2}. 

\section{Angular correlations}

\begin{figure} [tb]
\includegraphics[scale=0.5,clip]{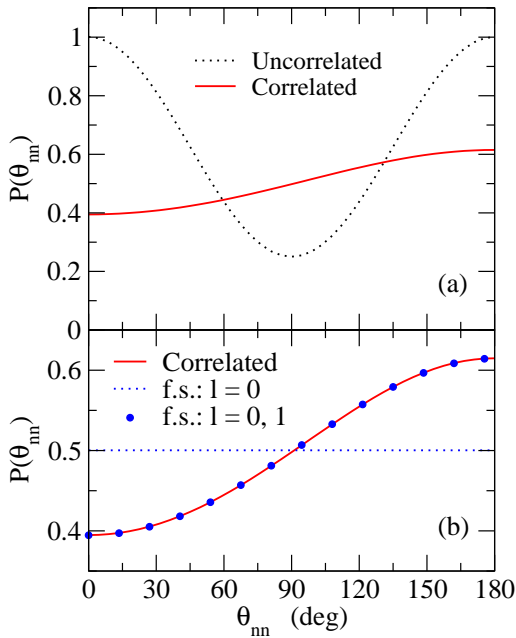}
\caption{(Color online) 
Upper panel: 
the angular correlations 
between the emitted neutrons from the two-neutron decay of $^{26}$O. 
The solid and the dotted lines show the correlated and the uncorrelated 
distributions, respectively. 
Lower panel: the decomposition of the 
correlated distribution into the angular momentum components. The dotted 
line and the filled circles are obtained by including the angular momentum of the 
final state up to $l$ = 0 and $l$ = 1, respectively. 
}
\label{fig:angular}
\end{figure}

We next discuss the angular correlation of the emitted neutrons from the ground state of 
$^{26}$O. 
The amplitude for emitting the 
two neutrons with spin components of $s_1$ and $s_2$ and momenta 
$\vec{k}_1$ and $\vec{k}_2$ is given by \cite{EB92,HS14}, 
\begin{eqnarray}
f_{s_1s_2}(\vec{k}_1,\vec{k}_2)&=&
\sum_{j,l}e^{-il\pi}e^{i(\delta_1+\delta_2)}\,
M_{j,l,k_1,k_2} \nonumber \\
&&\times \langle [{\cal Y}_{jl}(\hat{\vec{k}}_1)
{\cal Y}_{jl}(\hat{\vec{k}}_2)]^{(00)}|\chi_{s_1}\chi_{s_2}\rangle,
\label{angularamplitude}
\end{eqnarray}
where 
$\chi_s$ is the spin wave function and $\delta$ is the nuclear 
phase shift. 
$M_{j,l,k_1,k_2}$ is the decay amplitude given by 
\begin{equation}
M_{j,l,k_1,k_2}
=\langle (jj)^{(00)}|(1+vG_0)^{-1}|\Phi_{\rm ref}\rangle\,\sqrt{\frac{de_1}{dk_1}}\sqrt{\frac{de_2}{dk_2}}, 
\label{amplitude2}
\end{equation}
in which the unperturbed Green's function, $G_0$, 
is evaluated at $E=e_1+e_2$. 
The angular distribution is then obtained as 
\begin{equation}
P(\theta_{12})=4\pi\sum_{s_1,s_2}
\int dk_1dk_2\, |f_{s_1s_2}(k_1,\hat{\vec{k}}_1=0,k_2,
\hat{\vec{k}}_{2}=\theta_{12})|^2,
\label{angular}
\end{equation}
where we have set the $z$-axis to be parallel to $\vec{k}_1$ and 
evaluated the angular distribution as a function of the 
opening angle, $\theta_{12}$,  
of the two emitted neutrons. 
As in Ref. \cite{HS14}, for simplicity, 
we compute the correlated distribution 
only at the peak energy of the spectrum and then normalize the calculated distribution. 

The upper panel of Fig. \ref{fig:angular} shows the angular distributions thus obtained. 
In the absence of the correlation between the valence neutrons, 
the angular distribution is symmetric 
with respect to  $\theta_{12}=\pi/2$ (see the dotted line). 
On the other hand, in the presence of the interaction between the valence neutrons, 
the angular distribution 
becomes highly asymmetric, with an enhancement of the back-to-back emission \cite{HS14,GMZ13}, 
as is shown by the solid line. This is a natural consequence of 
the dineutron correlation in the momentum space shown in Fig. \ref{fig:density2-p}. 

As we have already discussed in Sec. II B, the dineutron correlation is caused by the interference between 
even- and odd- angular momentum configurations. 
In order to demonstrate this in connection to the angular correlation of the emitted neutrons, the lower 
panel of Fig. \ref{fig:angular} shows the decomposition of the distribution into several angular momentum 
components. 
The dotted line is obtained by including only $l$=0 in Eq. (\ref{angularamplitude}), which leads to a flat 
distribution reflecting the property of the $s$-wave.   
The filled circles, on the other hand, are obtained by including both $l$=0 and $l$=1. One can now see 
the enhancement of the back-to-back emission, due to the interference between the $l$=0 and $l$=1 components, 
reflecting the dineutron correlation. 

Moreover, one can also see that the angular distribution is almost exhausted only by the $l$=0 and $l$=1 components. 
One can view this as follows. 
That is, the original $d^2$ component in the reference 
wave function, $\Phi_{\rm ref}$, is scattered by the neutron-neutron 
interaction to the $s^2$ and $p^2$ configurations during the penetration of the centrifugal 
barrier, and are then observed as emitted neutrons from $^{26}$O \cite{GMZ13}. 
The operator $(1+vG_0)^{-1} = 1 -vG_0 +vG_0vG_0 -\cdots$ in Eq. (\ref{amplitude2}) has a responsibility 
for this multiple scattering process during the penetration. 
The mixing between the $s^2$ and $p^2$ configurations is such that 
the back-to-back emission is enhanced being consistent with the dineutron correlation. 
In principle, the two neutrons may be scattered 
to high partial waves, but these 
are suppressed due to high centrifugal barriers. 

\begin{figure} [tb]
\includegraphics[scale=0.5,clip]{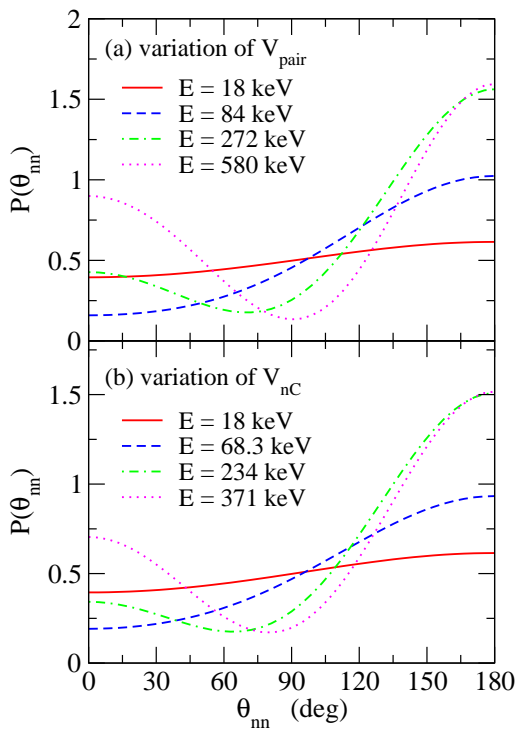}
\caption{(Color online) 
The dependence of the correlated 
angular distribution on the energy of the ground state. 
The upper panel is obtained by varying the energy by changing the pairing interaction between 
the two valence neutrons. On the other hand, the lower panel is obtained by shifting the resonance 
energy of the $d_{3/2}$ state in $^{25}$O keeping the strength of the pairing interaction to be the same. 
}
\label{fig:angular2}
\end{figure}

As compared to the angular distribution reported in Ref. \cite{HS14}, the degree of asymmetry is smaller in 
the present calculation shown in Fig. \ref{fig:angular}. This is due to the smaller ground state energy, that is, 
we employ the ground state energy of 18 keV while Ref. \cite{HS14} considered the energy of 140 keV. 
Since the energy is smaller, the $p$ wave contribution is smaller in the present calculation. This leads to a 
smaller admixture between the $s$-wave and the $p$-wave components, and thus the smaller asymmetry in the angular 
distribution. In order to see this, 
Fig. \ref{fig:angular2} shows the dependence of the angular distribution on the ground state 
energy. For this purpose, we vary the ground state energy in two different ways. 
In the first calculations, we vary it by changing the strength of the pairing interaction, 
while we vary the $d_{3/2}$ resonance energy of the $^{25}$O in the second calculation 
keeping the strength of the pairing interaction to be the same. In either way, one can see that the 
asymmetry indeed becomes larger as the ground state energy increases. 
If the ground state energy further increases, the $d$-wave starts contributing, and the forward angle components 
grow up, even though the back-to-back emission is still enhanced due to the dineutron correlation. 
  
Very recently, Kohley {\it et al.} extracted experimentally 
the three-body correlations from the ground state 
decay of $^{26}$O \cite{KBC15}. The observed angular correlation has shown 
an enhancement of the forward angle emissions, in contradictory to Fig. \ref{fig:angular}. 
Kohley {\it et al.} have also performed the Monte Carlo simulations and have figured out that the observed 
three-body correlations are insensitive to the theoretical predictions \cite{KBC15}. That is, even a theoretical 
calculation with the enhancement of back-to-back emission leads to the opposite trend 
in the actual measurement, {\it i.e.}, the enhancement of 
forward angles. Kohley {\it et al.} argue that this is due to the uncertainty in the momentum of the 
$^{24}$O nucleus at the reaction point in the target, which provides a large impact especially when 
the decay energy is small \cite{KBC15}. A confirmation of our prediction shown in Fig. \ref{fig:angular} therefore 
seems quite challenging at this moment, and a further experimental development will still be necessary.  

\section{Summary}

We have investigated the two-neutron emission decay of the unbound nucleus $^{26}$O using the 
$^{24}$O + $n$ + $n$ three-body model. To this end, we have calibrated the model parameters using 
the new experimental data measured at RIKEN. 
We have first discussed properties of the two-body subsystem, $^{25}$O = $^{24}$O + $n$. 
Using a Woods-Saxon potential which reproduces the energy of the $d_{3/2}$ resonance, we have shown that 
the calculated 
width of the $d_{3/2}$ resonance as a Gamow state 
agrees well with the experimental data. In addition to the known 
$d_{3/2}$ resonance, we have also found a broad $p_{3/2}$ and a relatively narrow $f_{7/2}$ 
resonance states above the $d_{3/2}$ resonance. 
We have then calculated the decay energy spectrum of $^{26}$O. The strength 
of the pairing interaction between the two valence neutrons was tuned in order to reproduce the ground 
state energy of the $^{26}$O nucleus. We have shown that this interaction leads to an excellent agreement 
with the experimental data for the excited 2$^+$ state. 
We have also investigated the ground state properties 
employing the bound state approximation. We have shown that the density distribution shows a clear signature 
of dineutron correlation, with an extended distribution of the valence neutrons. We have found that the 
ground state of $^{26}$O is dominated by the $(d_{3/2})^2$ configuration. In addition to the ground state, 
we have found at least two more excited 0$^+$ states, which consist mainly of the $(p_{3/2})^2$ and 
the $(f_{7/2})^2$ configurations, respectively, being consistent with the resonance structure of 
the two-body subsystem. For the $(f_{7/2})^2$ resonance, we have estimated also the resonance width. 
For the angular correlation of the two emitted neutrons, we have confirmed the result of our previous 
calculations, that is, an enhancement of the back-to-back emission. We have argued that this enhancement of 
the back-to-back emission is a clear signature of dineutron correlation in $^{26}$O, which can be understood 
in a simple term of uncertainty relation between the space and the momentum. 

The unbound nucleus $^{26}$O studied in this paper is a 
unique three-body system, which is unbound 
only slightly with an extremely small decay energy. 
This property would offer an interesting 
opportunity for further experimental and theoretical 
investigations on many-body correlations in neutron-rich nuclei. In this situation, the two-particle 
Green's function method will be a useful means in order to analyze experimental data. 
A theoretical challenge is to extend it to a 
four body decay, such as $^{28}$O. Another theoretical challenge 
is to apply it to the unbound $^{16}$Be and $^{13}$Li nuclei, both of which have been observed experimentally. 
To that end, a treatment of the deformation and the Borromean nature of the core nuclei would
 be crucial in the former 
and the latter unbound nuclei, respectively.

\section*{Acknowledgments}

We thank Y. Kondo, T. Nakamura, W. Nazarewicz, and B.A. Brown 
for useful discussions.

\end{document}